# STATUS OF CLIC MBQ AND DBQ PROTOTYPES PROCUREMENT

Michele Modena[1]

[1] CERN – TE Department, 1211 Geneva 23 – Switzerland

This paper gives the status of studies, design and prototype procurement for the magnets needed for the CLIC R&D and for the LAB and CLEX Test Programs.

## 1 Introduction

An intensive plan of tests on critical items of CLIC R&D project is actually planned at CERN starting from 2012. Among these tests it is planned to install in "LAB Test Program" and in the future CLIC Beam Test Facility CLEX several CLIC 2-beams Modules.

The CLIC 2-beams Modules are the backbone of the 2 Linacs constituting the major accelerator sub-system of CLIC. Status of Modules design and studies are presented in other contribution to this Workshop, here we will focalize on the status of studies, design and prototypes procurement for the magnets needed for the CLIC Modules Test Program.

The magnets present on the CLIC 2-beams Modules are of two families: Main Beam Quadrupoles (MBQ) and Drive Beam Quadrupoles (DBQ). Naming indicates clearly the purpose of these two type of magnets: MBQ are needed for the focalization of the two Main Beams; the beams that will collide into the IP at the center of the detector, while the DBQ are the quadrupoles needed to focalize the 48 Drive Beams needed to transfer the RF power to the Main Beams.

## 2 Status of Main Beam Quadrupoles design and procurement

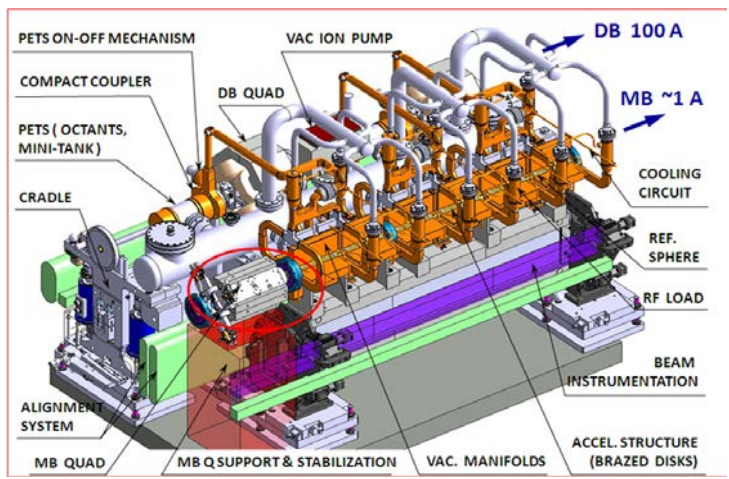

Figure 1: Design of a typical CLIC 2-beam Modules (Type1 in this specific case)

A typical 2-beam CLIC Module it is shown in Figure 1, where are clearly visible all the major components that must be assembled on the Modules. Among them, in the red oval, the MBQ (a Type1 MBQ in the case of Figure 1).

The MBQ are present in 4 different variants: Type 1, 2, 3, and 4. The only difference between the 4 variants is the length of the MBQ magnets. As mentioned the CLIC baseline include different modules layout. The Modules in the Tunnel will be of 5 types: Type 0 (without any MBQ) and Type 1 to 4 (containing each one a MBQ magnet of the same type of the module naming).

Figure 2 shows the layout of the different types of Modules. In total the 2 CLIC Linacs will contain ~ 4000 Modules with MBQ magnets.

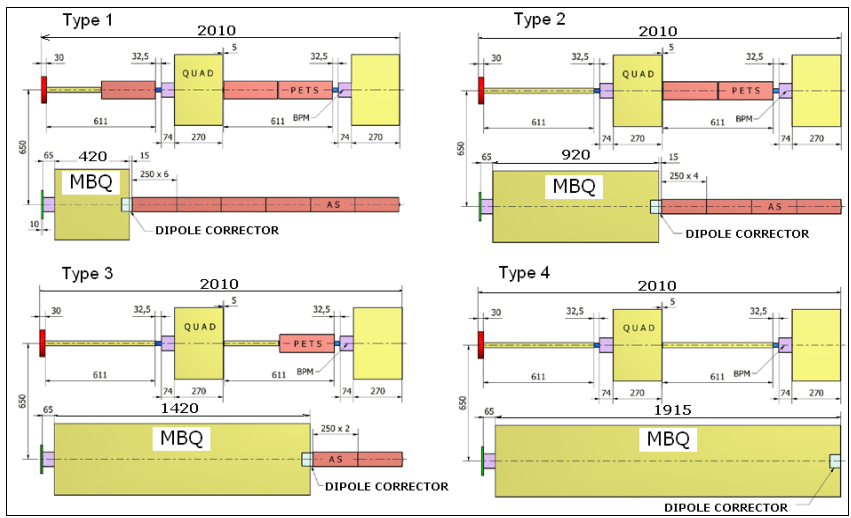

Figure 2: Layout of the four different types of Modules (Type 1, 2, 3, and 4) containing each one an MBQ of different length.

| Parameter | Value |
| --- | --- |
| Type | Type 1/2/3/4 |
| Quantity | 308/1276/964/1472 |
| Nominal field gradient | 200 T/m |
| Nominal integrated field | 70/170/270/370 T |
| Magnetic length | 350/850/1350/1850 mm |
| Maximum overall length | 420/920/1420/1915 mm |
| Magnet bore diameter | 10 mm |
| Good Field Region (GFR) | 4 mm |
| Integrated field gradient error | < 0.1% |

Table 1: Magnetic and geometric requirements for the MBQ magnets (4 variants)

Table 1 summarizes the geometrical and magnetic requirements for the MBQs. The magnets will provide a central field gradient of 200 T/m with a magnetic length of 350 mm, 850 mm, 1350 mm and 1850 mm for MBQ Types 1-4 respectively. The magnet bore diameter is 10 mm and the integrated field gradient quality has to be better than 0.1 % inside a Good Field Region (GFR) of 4 mm radius. The restrictions imposed by the modules layout, limit the maximum overall length of the magnets to 420 mm, 920 mm, 1420 mm and 1915 mm for MBQ Type 1, 2, 3 and 4 respectively.

## 2.1 MBQ Design

In 2009 the design and procurement of MBQ Type 1 and Type 4 prototypes was launched. The priorities of the MBQ R&D were to develop a design for:
a) a compact magnet (it needs to be actively stabilized so the weight must be minimized)
b) the development of a "simple" magnet design (investigating the optimization of the tolerances, cost, assembly procedures)
c) a procurement for the total number of prototypes needed for the CLIC test program with and without beam (a maximum of seven Modules is planned).

The global view of the MBQ Type 4 prototype is shown in Figure 4. Due to priorities mentioned at previous point b), a common design (cross-section) is used for all types of the MBQ. The magnet yoke consists of four pieces (quadrants) made of solid steel blocks, which permit installation of the coils around each pole. Low carbon steel AISI 1010 was chosen as a yoke material for reason of cost, availability and ease of machining.

The water cooled coils of 17 turns each, are made of hollow copper conductor with a square cross section of 5.6 mm × 5.6 mm, edge rounding of 1.0 mm and a circular cooling hole with a diameter of 3.6 mm. The main parameters of the MBQ Type1 and Type 4 models are listed in Table 2.

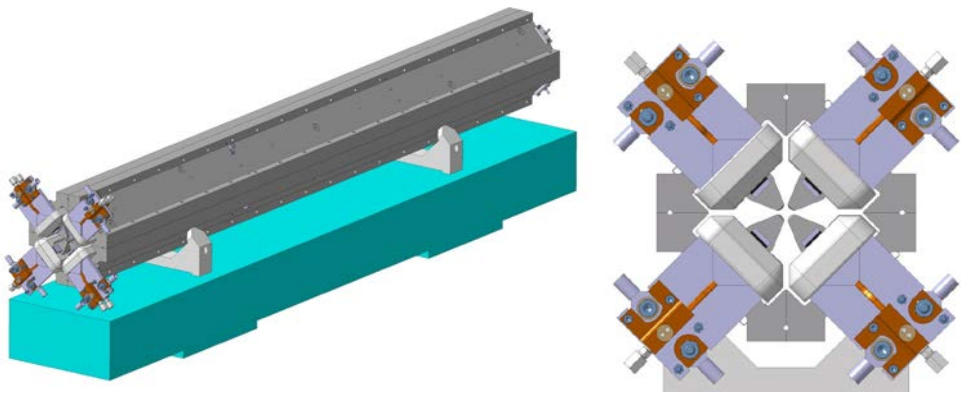

Figure 4: Views of the Type4 MBQ final design.

| Parameter | Type 1 | Type 4 | Unit | Parameter | Type 1 | Type 4 | Unit |
|---|---|---|---|---|---|---|---|
| Operating gradient | 210 | 202.2 | T/m | Windings per pole | 17 | 17 | |
| Integrated gradient | 70 | 370 | T | Conductor dimensions | 5.6×5.6, Ø=2.5 | | mm |
| Magnetic length | 333.8 | 1829.4 | mm | Total resistance | 48.2 | 220.3 | mΩ |
| Yoke length | 332 | 1827 | mm | Total inductance | 7.2 | 39.6 | mH |
| Total length | 420 | 1915 | mm | Power | 0.95 | 3.61 | kW |
| Magnet bore diameter | 10 | 10 | mm | Cooling circuits/magnet | 1 | 4 | |
| Nominal current | 140 | 128 | A | Pressure drop | 5.9 | 6.3 | bar |
| Current density | 6.8 | 6.3 | A/mm$^2$ | Temperature rise | 15 | 15 | K |

Table 2: Main characteristics of the MBQ magnets (Type1 and Type4 variant)

## 2.2 MBQ Prototypes Procurement

Two magnets, one Type4 and one Type1, were assembled, fully tested and are now delivered to colleagues for CLIC Stabilization studies and LAB Test Program. Photos of the 2 assembled prototypes are shown in Figure 5 and 6.

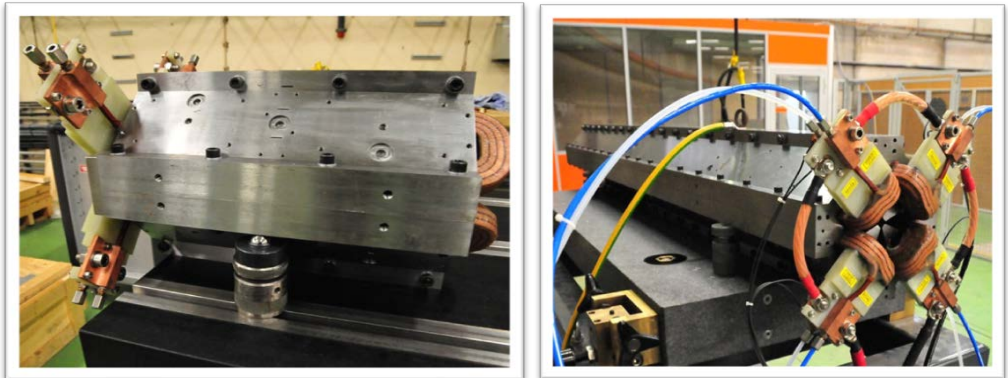

Figure 5-6: Photos of the assembled Type1 and Type4 MBQ prototypes

Coils for other two magnets are procured and are at CERN and we are now selecting other Manufacturers for the iron quadrants, since we had an unlucky experience with the first quadrants set procurement.

The procurement of the coils was straightforward, but we have experienced difficulties in the procurement of the iron quadrants. Despite the machining quality required and promised by the Manufacturer, the quadrants manufacturing was out from the specified tolerances (0.010 mm for Type 1 and 0.020 mm for Type 4 quadrants). In the best cases the critical surfaces (ex. the pole profiles) have a "relative" good tolerance but the surfaces are shifted up to 0.6 mm respect to their nominal position. Cases like that are evident manufacturing errors that are not representative of the best possible achievable machining that we were trying to investigate.

Figure 7 shows a typical metrology report (from CERN Metrology Lab) for a Type 1

quadrant. The pole profile (hyperbolic, tangents and tips) is evidently out of tolerance. The tolerances (± 0.010 mm) are also indicated in the drawing.

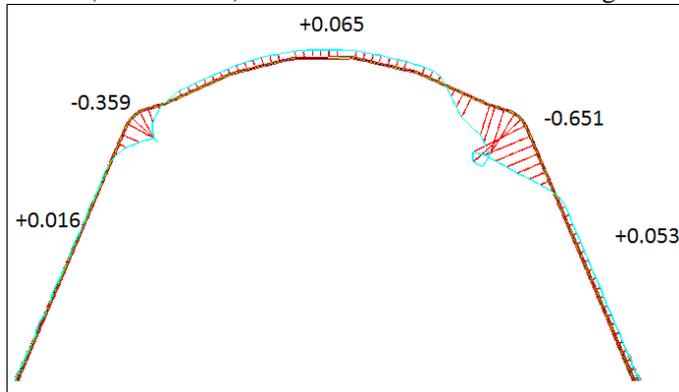

Figure 7: Metrological report of a MBQ quadrant pole profile

### 2.3  MBQ Prototype Magnetic Measurement

The integrated gradient fo the MBQ prototypes was measured with the SSW (Single Stretched Wire) technique.

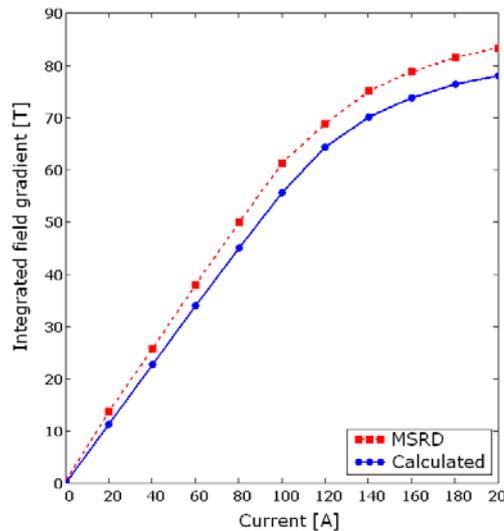
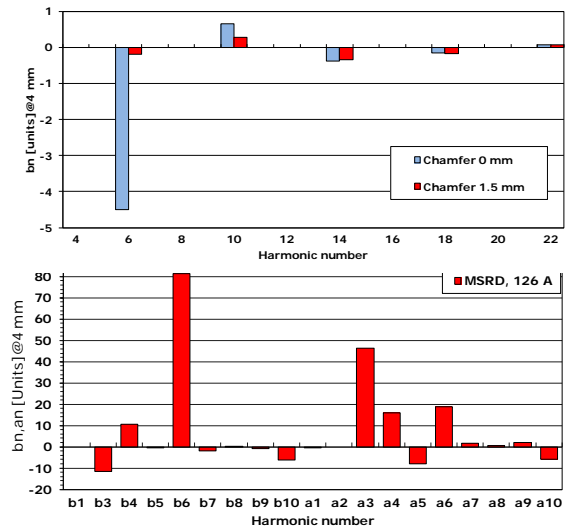

Figure 8: excitation curve of the MBQ Type 1 prototype.
Figure 9: Computed Field Quality for the Type 1 MBQ (top)
Figure 10: Measured Field Quality for the Type 1 MBQ (botton)

In Figure 8 is shown the excitation curve of the Type 1 MBQ prototype. The measured gradients are slightly higher of the computed one due to a slightly better permeability of the steel utilized compared to the one used in the computation and also because of a small reduction (out of tolerance) of the quadrupole inner bore.

The expected (computed) Field Quality (expressed in multipoles content) is shown in Figure 9. It is evident the effect of the chamfering of the poles extremities in order to reduce mainly the dodecapole component (b6). Only the "permitted" multipoles are present, being this the result of a computation for a "perfectly" manufactured and assembled quadrupole.

The measured multipoles (by Rotated Vibrating Wire, a new technique developed at CERN [2]) are shown in Figure 10. It is evident how the mentioned not acceptable quality of the machined quadrants is translated in huge multipoles content.

It must be mentioned that the Rotated Vibrating Wire method is today not guaranteeing an high accuracy mainly for b6 (intended as the precision of the multipoles measurements respect to a "true value" measured with other means). The total error (accuracy + repeatability) is estimated to be ± 10 units.

A better knowledge of the field quality will be also come by a new miniaturized Rotating Coils measurement system actually under commissioning at CERN.

## 2.4 MBQ Version 2 Prototype procurement

In order to complete the MBQ procurement list (four magnets in total: 2 Type 1 and 2 Type4) we are now proceeding with the procurement of 2 more magnets.

In the aim of an easier machining and assembly and an improved field quality, the quadrants of the next magnets will be slightly modified respect to the previous design (Version 1) of MBQ. The cross section and pole profile for the Version 2 design are shown in Figure 14 and 15.

It can be noticed, respect to Version 1, as some iron was added externally in order to have simpler and bigger straight reference surfaces. As a consequence, weight of the quadrants will increase of ~20%. The other change is in the pole profile previously composed by hyperbolic + tangents + tips arcs, and now composed only by circles arcs (+ tangents at the extremities of the pole). Coils will be identical to the ones of Version 1.

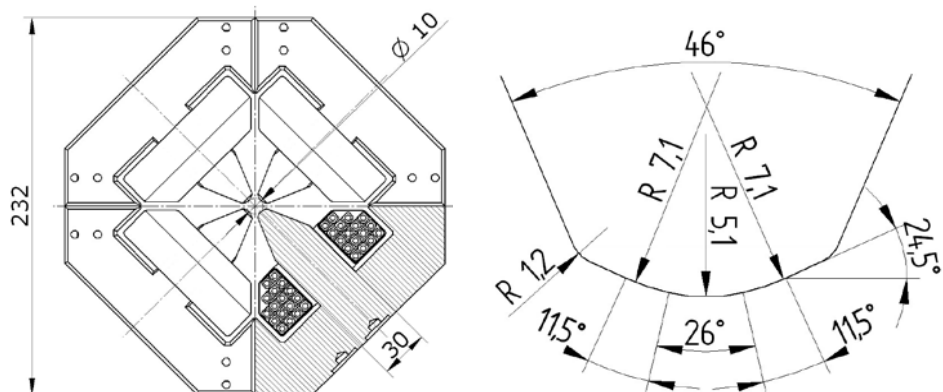

Figure 11 and 12: Cross-section and pole profile for the MBQ Version 2 now under procurement.

## 3 Status of Drive Beam Quadrupoles design and procurement

Drive Beam quadrupole are the most populated magnet family in CLIC Complex. As mentioned before they will provide the focalization of the Drive Beam. About 40000 units would be needed in the CLIC 3TeV layout.

The position of the DBQ on the CLIC 2-beams Modules is indicated in Figure 13 by the two red ovals: each CLIC Modules (Type 0, 1, 2, 3,and 4) will contain 2 DBQ magnets.

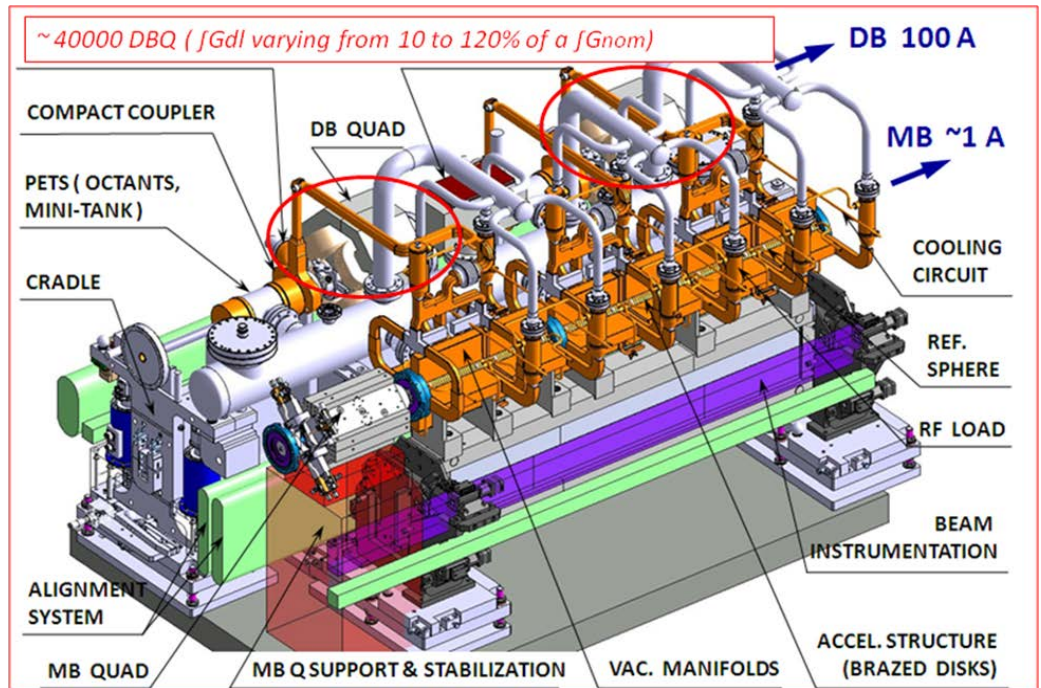

Figure 13: DBQ (in the red ovals) on the CLIC 2-beams Modules

Characteristic of DBQs magnets is the gradient required; the gradient will be maximum at the beginning of the Decelerator (starting of the Linac) and minimum at the end. Gradient variation (including some margin for studies and commissioning) will be between 10 and 120% of a nominal maximum value of 62.6 T/m. The space allocated on the Modules for the DBQ is constant all along the Decelerator (longitudinally: only 286 mm per magnet are available). With these tight boundary conditions the design of a classical electromagnetic DBQ is quite challenging (space, power consumption, etc.). Anyway a solution was found and 8 prototypes, needed for CLEX the CLIC Test Beam Facility, are now under procurement. Table 3 shows the excitation curve and the main parameters for the DBQ EM design. Figure 14 shows a global view of the proposed design of the 8 prototypes. This design is under check for integration on the CLIC 2-beams Module prototypes.

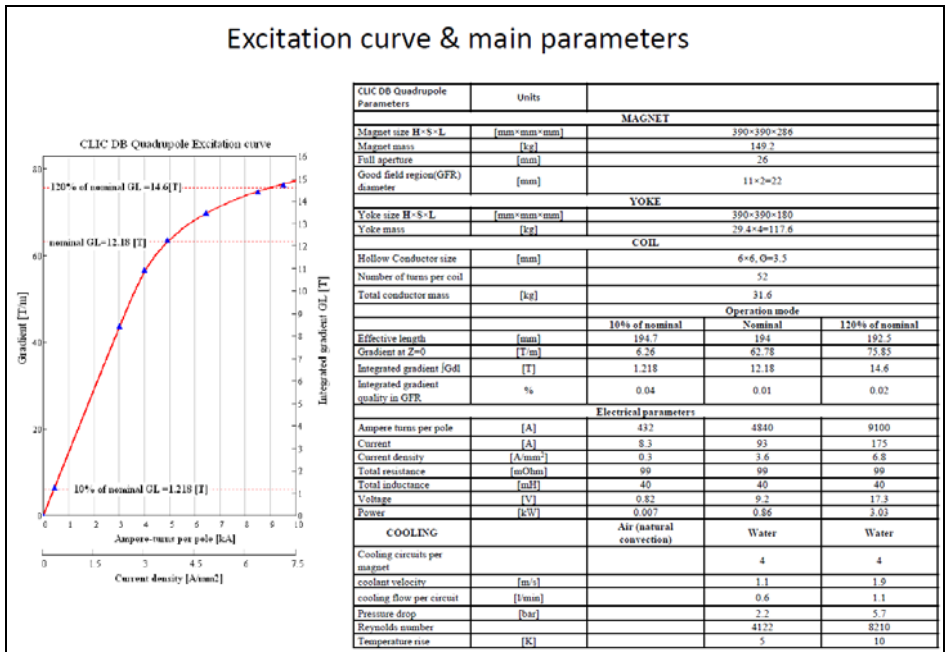
Table 3: DBQ excitation curve and main parameters.

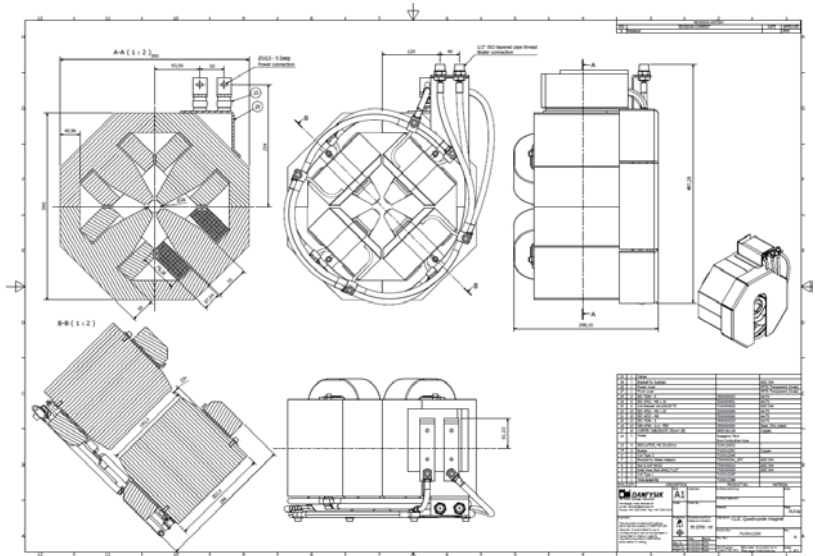
Figure 14: DBQ (in the red ovals) on the CLIC 2-beams Modules

Since the power consumption of the DBQ magnets placed in the first part of the Decelerator is relatively quite high due to iron saturation (this is a consequence of the small space available on the modules that unfortunately is not possible to increase today), other solutions for the DBQ magnet design are under study.

Colleagues of Cockcroft Institute (UK) are developing a quadrupole design based on tunable permanent magnet technology [3]. In order to cover the full range of gradient tunability required, it will be necessary to develop two different design.

The first one (for the higher Gradient requirement range) is now completed and a first prototype is under procurement. This prototype will be accurately measured and tested at CERN. The 2$^{nd}$ prototype (for the lower Gradient range) is expected for 2013 and will be also compatible with CLEX Test Facility Modules making possible to test it with beam. A global view of the 1$^{st}$ prototype is shown in Figure15.

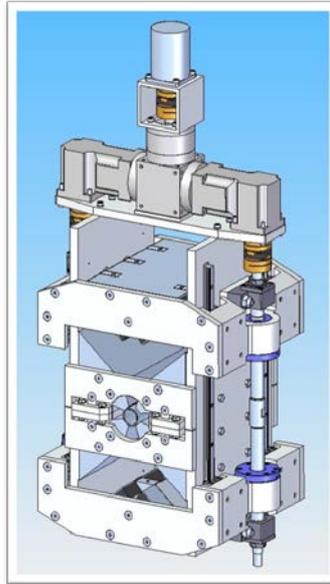

Figure 15: View of the alternative design of a DBQ based on tunable permanent magnets.

## 4 Acknowledgements

This work is based on the activities of many person of CERN in particular: R. Leuxe, A. Newborough, E. Solodko, M. Struik, A.Vorozhtsov.

Thanks to D.Tommasini for the support of the activities and to the CERN-BE colleagues E. Adli, D. Schulte, G. Sterbini for the continuous discussion on the magnet requirements.